\documentclass[12pt]{article}
\usepackage{amsfonts,graphicx}
\newcommand{\beq}{\begin{eqnarray}}
\newcommand{\eeq}{\end{eqnarray}}
\newcommand{\beqa}{\begin{eqnarray}}
\newcommand{\eeqa}{\end{eqnarray}}
\newcommand{\beqar}{\begin{eqnarray*}}
\newcommand{\eeqar}{\end{eqnarray*}}

\newcommand{\eg}{{\it e.g.,}\ }
\newcommand{\ie}{{\it i.e.,}\ }

\newcommand{\norm}[1]{\raise.3ex\hbox{:}#1\raise.3ex\hbox{:}}

\newcommand{\labell}[1]{\label{#1}} %\qquad_{#1}} %{\label{#1}}
\newcommand{\reef}[1]{(\ref{#1})}
\newcommand{\Tr}{{\rm Tr}}
\newcommand{\STr}{{\rm STr}}

\newcommand\hi{{\rm i}}

\newcommand\ls{\ell_s}

\newcommand\tF{{\tilde F}}
\newcommand\tB{{\tilde B}}

\newcommand\dtF{{{^*\!\tilde F}}}
\newcommand\vB{\vec{B}}
\newcommand\nab{\vec{\nabla}}
\newcommand\identity{{\mathbf{1}}}
\newcommand\bbR{{\mathbb{R}}}

\begin{document}

\thispagestyle{empty}
\rightline{\small hep-th/0105035 \hfill McGill/01-06}
\vspace*{2cm}

\begin{center}
{\bf \Large Fuzzy Funnels: Non-abelian Brane Intersections} 
\footnote{Based on the talk presented by R. C. Myers at 
Strings 2001, Bombay, India}
\vspace*{1cm}

Neil R. Constable\footnote{E-mail: constabl@hep.physics.mcgill.ca},
Robert C. Myers\footnote{E-mail: rcm@hep.physics.mcgill.ca}
and \O yvind Tafjord\footnote{E-mail: tafjord@physics.mcgill.ca}

{\it Department of Physics, McGill University}\\
{\it Montr\'eal, QC, H3A 2T8, Canada}\\
\vspace{2cm} ABSTRACT
\end{center}
We discuss dual formulations of D-brane intersections. The duality
is between world volume field theories of different dimensionalities
which both describe the same D-brane configuration but are valid in 
complementary regions of parameter space. We discuss the duality in terms
of bion configurations involving D-strings orthogonally intersecting 
both D3-branes and D5-branes.

\vfill \setcounter{page}{0} \setcounter{footnote}{0}
\newpage

\section{Introduction}

In string theory we have become accustomed to the notion that many
physical systems possess several equivalent or rather dual mathematical
descriptions which may be distinguished by their usefulness 
in different regions of parameter space. The S-,T-,U-dualities and
AdS/CFT are perhaps the most well known and useful of these relations. 
Of these AdS/CFT~\cite{juan1} is a remarkable example of a duality relating 
theories in different space-time dimensions. In this case the duality is
holographic in the sense that a theory of gravity living in an AdS space is
completely described by a quantum field theory living on the ``boundary'' of the
AdS space. Holography however need not be a pre-requisite for such 
relationships between theories in different dimensions. 
Here we will discuss two simpler examples of systems which 
have alternate descriptions as field theories in different dimensions.
First we consider the well known example of D1-branes orthogonally 
intersecting a D3-brane. In the low-energy world volume theory
of the D3-brane this is described as a magnetic monopole plus a 
non-trivial profile for one of the transverse scalar fields in the 
($3+1$)-dimensional Born-Infeld action\cite{calmaldetc}. If one is interested
in the dynamics of this system far away from the core of the monopole
then one simply expands the Born-Infeld action around the monopole-scalar 
background. If on the other hand one wishes to probe this system near
the point at which the D1-branes are attached to the D3-brane, \ie
near the core of the magnetic monopole, the low energy 
theory on the D3-brane is no longer valid as the fields are rapidly
varying on the string scale
at this point. Fortunately this D-brane intersection has another description 
in terms of the ($1+1$)-dimensional theory living on the 
D1-branes\cite{ded,bioncore}. Here the D1-brane system is found to have
non-commutative solutions in which three of the transverse scalar fields 
open up into a fuzzy funnel (so called because the cross sections
are fuzzy two-spheres) as one moves along the D1-branes. This funnel turns
out to be a source for the correct Ramond-Ramond (RR) field to be identified
with an orthogonal D3-brane\cite{bioncore}. Since the ($1+1$)-dimensional
Born-Infeld action can be trusted when the radius of the funnel is small,
this formulation provides an accurate description of the physics
near the point of attachment, or rather at the monopole core.
Thus the ($3+1$)-dimensional theory on the D3-branes
and the ($1+1$)-dimensional theory on the D-strings provide dual
frameworks with which to describe the D3$\perp$D1 intersection. In either
case, the low energy world volume theory cannot provide a complete
account of this system, but what is remarkable is the degree to
which the two descriptions agree in the limit of large magnetic
charge\cite{bioncore}. 

Another example of this type of duality is provided by the orthogonal
intersection of D1-branes and D5-branes recently examined in
ref.~\cite{nonabel}.
This system is very similar to the D3$\perp$D1-system in that
the theory on the world volume of the D5-branes involves a non-trivial
gauge field configuration as well as a scalar field profile giving rise to
a bion spike configuration. Again, near the core of the spike the
($5+1$)-dimensional
Born-Infeld action is no longer trustworthy and we must revert 
to the dual description provided by the D1-branes. As in the D3-brane case an
orthogonal D5-brane can be realised as a non-commutative funnel involving five
of the transverse scalars with a cross section given by fuzzy
four-spheres. The configuration carries the correct charge to be identified
with a D5-brane and as before provides an accurate description of the 
physics near the point at which the D1-branes are attached to the the 
D5-branes.

The remainder of this note is organized as follows. In section 2, we
review the dual formulations of the D3$\perp$D1-intersections as presented
in ref.~\cite{bioncore}. In section 3, we provide a similar discussion
of the the D5$\perp$D1 system, as given in ref.~\cite{nonabel}. In section 4,
we conclude with a discussion of further applications of these dualities and
some open questions.

\section{Dual Formulations of D3$\perp$D1 Intersections}
It was originally recognized in ref.~\cite{calmaldetc} that the world
volume theory of a D-brane supports ``solitonic'' configurations corresponding
lower-dimensional branes protruding from the original D-brane. In particular,
in the context of D3-branes, BPS monopoles correspond to an orthogonal
D-string. This system has been extensively studied in the
literature\cite{waves,K/T,larus} and we review some of
the salient results here.

The low energy dynamics of 
a (single) D3-brane in a Minkowski space background is described
by the Born-Infeld action\cite{bin},
\beq
S=\int dt\, {\mathcal L}=-T_3\int d^4\sigma\sqrt{-\det\left(\eta_{ab}+
\lambda^2\partial_a\phi^i\partial_b\phi^i+\lambda F_{ab}\right)},
\labell{act3}
\eeq
where $\lambda\equiv2\pi\alpha'=2\pi\ls^2$. Here
we have implicitly used static gauge, with $\sigma^a\,(a=0,\ldots,3)$ denoting
the world-volume coordinates while $\phi^i\,(i=4,\ldots,9)$ are the
scalars describing transverse fluctuations of the brane. $F_{ab}$ is the
field strength of the $U(1)$ gauge field on the brane.

General considerations of intersecting branes\cite{strominger} indicate
that a D-string ending on a D3-brane should act as a source of the first
Chern class of the world-volume gauge field. Hence, as
mentioned above, D-strings appear as BPS magnetic monopoles within the
three-brane world volume\cite{calmaldetc}. The solution involves
exciting one of the scalar
fields, say $\phi\equiv\phi^9$, as well as a magnetic field
$B^r={1\over2}\epsilon^{rst}F_{st}\ (r,s,t=1,\ldots,3)$. One can show
that it is consistent with the equations of motion to
set all the other fields equal to zero. For static configurations, we then
evaluate the energy,
\beqa
E=-{\mathcal L}&=&T_3\int d^3\sigma\sqrt{\lambda^2|\nab\phi\mp\vB|^2+
(1\pm\lambda^2\vB\cdot\nab\phi)^2}\nonumber\\
&\ge&T_3\int d^3\sigma\left(1\pm\lambda^2\vB\cdot\nab\phi\right).
\label{enineq}
\eeqa
The first term in this lower bound is simply the energy of the D3-brane.
The second term
is a total derivative: $\vB\cdot\nab\phi=\nab\cdot(\vB\phi)$ using
the Bianchi identity $\nab\cdot\vB=0$. Hence this term is topological,
depending only on the boundary values of the fields specified at infinity
and near singular points (after introducing a cutoff, such as near $r=0$ below).
Therefore, the last line of eq.~(\ref{enineq}) provides a true minimum 
of the energy for a given set of boundary conditions\cite{gauntlett}. 

The lower bound in eq.~\reef{enineq} is achieved when
\beq
\nab\phi=\pm\vB\ ,
\labell{ssusy}
\eeq
which coincides with the BPS condition for the magnetic
monopoles\cite{calmaldetc}. Further, using the Bianchi identity,
eq.~\reef{ssusy}
implies $\nab^2\phi=0$ and so we are looking for harmonic functions
on the D3-brane. The simplest solution, corresponding to the ``bion spike,''
is given by
\beq
\phi={N\over 2r}\ ,\qquad\qquad
\vB=\mp{N\over 2r^3}\vec{r}\ ,
\label{phisol}
\eeq
where $r^2=(\sigma^1)^2+(\sigma^2)^2+(\sigma^3)^2$, and
$N$ is an integer due to quantization of the magnetic charge. The
energy of this configuration is readily computed to be
\beq\labell{d3d1energy}
E=T_3\int d^3\sigma+N T_1\int_0^\infty d(\lambda\phi),
\eeq
where $T_1=(2\pi \ls)^2 T_3$. Here we note that $\lambda\phi$ is
the physical distance in the transverse direction.
Hence we have recovered the energy expected for a BPS
configuration consisting of $N$ semi-infinite D-strings ending
on an orthogonal D3-brane.
Other considerations, such as the effective charge
distribution and fluctuations on the strings\cite{waves,K/T}, further
confirm that the bion spike describes the D3$\perp$D1-system to a good
approximation.

The D3$\perp$D1-system
has a dual description in the D-string theory, which has also
received a great deal of attention in the literature\cite{ded,
gauntlett,morefun}. As emphasized in ref.~\cite{bioncore}, interpreting
the D-string world volume solutions in terms of
noncommutative geometry completes the correspondence to the results
in D3-brane theory. The low energy dynamics of $N$
D-strings in a flat background is well described by the
non-abelian Born-Infeld action\cite{dielec,yet2}:
\beq
S=-T_1\int d^2\sigma\, \STr\sqrt{-\det\left(\eta_{ab}+
\lambda^2\partial_a\Phi^i\, Q^{-1}_{ij}\,\partial_b\Phi^j\right)
\ \det\left(Q^{ij}\right)}\ ,\labell{action}
\eeq
where $Q^{ij}=\delta^{ij}+i\lambda[\Phi^i,\Phi^j].$
Again we are assuming static gauge where the two worldsheet coordinates are
identified with $\tau=x^0$ and $\sigma=x^9$. The transverse scalars
$\Phi^i\ (i=1,\ldots,8)$ are now $N\times N$ matrices transforming
in the adjoint representation of the $U(N)$ worldsheet gauge symmetry.
The symmetrized trace prescription\cite{yet}, denoted by $\STr$, 
requires that we symmetrize over all permutations of $\partial_a\Phi^i$ and
$[\Phi^i,\Phi^j]$ within the gauge trace upon expanding the square root.

To find static solutions describing the D3$\perp$D1-system, we allow three of
the transverse scalars, $\Phi^i$ ($i=1,2,3$), to depend on the spatial
coordinate $\sigma$. Evaluating the determinants in eq.~\reef{action}
the energy becomes
\beqa
E&=&T_1\int d\sigma\,\STr\sqrt{\lambda^2(\partial_\sigma\Phi^i
\mp{i\over2}\epsilon^{ijk}[\Phi^j,\Phi^k])^2+
(1\pm{i\over2}\lambda^2\epsilon^{ijk}\partial_\sigma\Phi^i
[\Phi^j,\Phi^k])^2}\nonumber\\
&\ge&T_1\int d\sigma\,\STr\left(1\pm{i\over2}\lambda^2\epsilon^{ijk}
\partial_\sigma\Phi^i
[\Phi^j,\Phi^k]\right)\nonumber\\
&=&NT_1\int d\sigma\pm{i\over3}\lambda^2 T_1\,\int d\sigma\,
\partial_\sigma\!\Tr\left(\epsilon^{ijk}\Phi^i\Phi^j\Phi^k\right).
\labell{d1d3enbound}
\eeqa
Here the lower
bound is again the sum of a trivial term (the energy of the $N$
D-strings) and a topological term. The minimum energy condition is
\beq
\partial_\sigma\Phi^i=\pm{i\over2}\epsilon^{ijk}[\Phi^j,\Phi^k],
\labell{nahm}
\eeq
which can be identified as the Nahm equations\cite{ded,nahm}. 
The solution dual to a single-centered BPS monopole is given by
\beq
\Phi^i(\sigma)=\pm{\alpha^i\over2\sigma},
\labell{Phisol}
\eeq
where the $\alpha^i$ are an $N\times N$ representation of the $SU(2)$
algebra: $[\alpha^i,\alpha^j]=2i\epsilon^{ijk}\alpha^k$.
The Casimir $C$ for this algebra is defined by
$\sum (\alpha^i)^2= C\,\identity_N$, where $\identity_N$ the $N\times N$
identity matrix. We will focus on the
irreducible $N\times N$ representation 
for which $C=N^2-1$. This noncommutative scalar field configuration
(\ref{Phisol}) describes a fuzzy two-sphere\cite{two} with a physical radius
\beq
R(\sigma)=\lambda\sqrt{\Tr[\Phi^i(\sigma)^2]/N}=
{\sqrt{C}\pi \ls^2\over\sigma}
={N\pi\ls^2\over\sigma}\sqrt{1-1/N^2}.
\labell{radstr}
\eeq
Hence the solution describes a ``fuzzy funnel'' in which the D-strings
expand to span the $X^{1,2,3}$ hyperplane at $\sigma=0$.
This geometry can be compared to the D3-brane solution (\ref{phisol}) after
relabeling $\sigma\rightarrow\lambda\phi$ and $R\rightarrow r$. We see that
both descriptions yield the same geometry in the limit of large $N$,
up to $1/N^2$ corrections. There is similar agreement for other quantities.
The energy can be evaluated from the boundary term in eqn.~\reef{d1d3enbound} 
and is found to be,
\beq
E=N T_1\int_0^\infty d\sigma+(1-1/N^2)^{-1/2}T_3\int 4\pi R^2 dR,
\labell{d1d3energy}
\eeq
while the non-abelian Wess-Zumino couplings of ref.~\cite{dielec} include,
\beq
i\lambda\mu_1\int\Tr P[{\rm{i}}_{\Phi}{\rm{i}}_{\Phi}C^{(4)}]=
\mp i\mu_3(1-1/N^2)^{-1/2}\int dt\,4\pi R^2dR\,C^{(4)}_{t123} 
\eeq
which indicates that, upto $1/N^2$ corrections, the funnel is a source 
for precisely the correct RR four-form
field to be identified with a D3-brane. Hence at least in the large $N$ limit, 
the two dual descriptions are in good agreement.

Note, however, that the Born-Infeld actions, eqs.~\reef{act3}
and \reef{action}, only describe low energy dynamics and do not
provide a complete theory in either case. Therefore, in general, one must
regard these two dual descriptions as complimentary. That is, the D-string
theory gives reliable results for small $r$ near the core of the spike, while
the D3-brane theory is valid for $r$ large. However, this makes the
strong agreement
between the two dual descriptions seem remarkable, so one must
be more precise about the range of validity of each theory\cite{bioncore}.
The Born-Infeld action does not include higher derivative stringy corrections,
but one finds that these are negligible for the D3-brane bion when
$R\gg\ls$ and for the D-string funnel when $R\ll N\ls$. The nonabelian
D-string action \reef{action} also neglects certain higher commutator
corrections\cite{hicom}, and one can argue that their contributions remain
small for $R\ll \sqrt{N}\ls$. Hence for large $N$, there is a large interval
in which both low energy actions give a reliable account of the
D3$\perp$D1-intersections. This overlap is the source of the strong
agreement between the two descriptions in this large $N$ regime.

\section{Dual Formulations of D5$\perp$D1 Intersections}
 
Given that the fuzzy geometry encountered in eq.~\reef{Phisol} was put in
by hand it is natural to consider
generalizations involving fuzzy geometries other than the two-sphere. In this
section, 
following ref.~\cite{nonabel}, we will consider the D-string theory in which
five  transverse scalars are excited and hence extend the above analysis by
considering fuzzy four-spheres. 

Our starting point is again the low energy action for $N$ D-strings,
however, we now consider static configurations involving
five (rather than three) nontrivial scalars, 
$\Phi^i$ with $i=1,\ldots,5$. In this case, the action (\ref{action}) becomes
\beqa
S&=&-T_1\int d^2\sigma\,\STr\left\{
1+\lambda^2(\partial_\sigma\Phi^i)^2+2\lambda^2\Phi^{ij}\Phi^{ji}
+2\lambda^4(\Phi^{ij}\Phi^{ji})^2-4\lambda^4\Phi^{ij}\Phi^{jk}
\Phi^{kl}\Phi^{li}+\phantom{1\over4}\right.\nonumber\\
&&\!\!\left.
\phantom{1}+2\lambda^4(\partial_\sigma\Phi^i)^2\Phi^{jk}
\Phi^{kj}-4\lambda^4\partial_\sigma\Phi^i\Phi^{ij}\Phi^{jk}
\partial_\sigma\Phi^{k}+{\lambda^6\over4}(\epsilon^{ijklm}
\partial_\sigma\Phi^i\Phi^{jk}\Phi^{lm})^2\right\}^{1/2},
\labell{expact}
\eeqa
where we have introduced the convenient notation,
$\Phi^{ij}\equiv{1\over2}\left[\Phi^i,\Phi^j\right]$.

To construct a new funnel solution, we consider the following ansatz:
\beq
\Phi^i(\sigma)=\pm\frac{R(\sigma)}{\sqrt{c}\lambda}\,G^i,\ \ i=1,\ldots,5\ ,
\labell{ansatz}
\eeq
where $R(\sigma)$ is the radial profile
and $G^i$ are the matrices constructed in ref.~\cite{wati4}
--- see also ref.~\cite{grosse}. The definition and many useful
properties of the $G^i$ matrices may be found in refs.~\cite{wati4,nonabel}. 
Here, we simply note that
the $G^i$ are given by the totally symmetric $n$-fold tensor product
of $4\times4$ gamma matrices, and that the dimension of the matrices
is related to the integer $n$ by
\beq
N={(n+1)(n+2)(n+3)\over6}. \labell{Nnrel}
\eeq
The solution will describe a funnel whose cross-section is a
fuzzy four-sphere with a physical radius
$R(\sigma)=\lambda\sqrt{\Tr[\Phi^i(\sigma)^2]/N}$.
The latter identification requires choosing the normalization
constant $c$ to be the ``Casimir'' associated with the $G^i$
matrices, \ie $G^i G^i=c\,\identity_N$, which is given by $c=n(n+4)$.

An immediate puzzle is that to leading order in $\lambda$ the equations of motion for this
system are exactly the same as for the case of three scalars in the previous section and
therefore the funnel profile will again be $R\sim \sigma^{-1}$. In fact,
this behavior is universal to any D-string funnel in the small $R$
regime\cite{bioncore}. If however the present configuration is
to represent a D5-brane then one would anticipate that the funnel would follow a 
profile given by $R\sim \sigma^{-1/3}$ appropriate for a harmonic function in five spatial 
directions. As we now show the resolution of this puzzle is that the higher 
order terms appearing in the action~\reef{expact} effect the necessary transition. 

Since dealing with the full equations of motion following from this action is tedious and
largely unenlightening we will not consider them here. We do note however that inserting the 
ansatz~\reef{ansatz} into these equations yields a single differential equation for 
${R}(\sigma)$. Knowing this we simply insert our ansatz into the action and making
use of the identities satisfied by the $G^{i}$ obtain an action for the radial profile,
\beq
S=-NT_1\int d^2\sigma\sqrt{1+(R')^2}[1+4R^4/(c\lambda^2)],
\label{Raction}
\eeq
which then in analogy with eqs.~\reef{enineq} and \reef{d1d3enbound}
yields the following bound on the energy,
\beqa
E
%&=&NT_1\int d\sigma\left[\left(R'{\mp}\sqrt{8R^4/(c\lambda^2){+}16R^8/
%(c\lambda^2)^2}\right)^2+
%\left(1{\pm} R'\sqrt{8R^4/(c\lambda^2){+}16R^8/(c\lambda^2)^2}\right)^2
%\right]^{1/2}\nonumber\\
&\ge&NT_1\int d\sigma
\left(1\pm R'\sqrt{8R^4/(c\lambda^2)+16R^8/(c\lambda^2)^2}\right).
\labell{Renergy}
\eeqa
This is again a sum of the trivial term and a topological term.
The equality is saturated when
\beq
R'=\mp\sqrt{8R^4/(c\lambda^2)+16R^8/
(c\lambda^2)^2}.
\labell{Rpeq}
\eeq
Note that this equation is also compatible with the full equation of motion.
We may write the solutions of eq.~\reef{Rpeq}
in terms of elliptic functions, however, the
geometry is more clearly exhibited by considering
various limits. For small $R$,  the $R^4$ term under the square
root dominates, and we find the funnel solution 
\beq
R(\sigma)\simeq{\sqrt{c}\lambda\over2\sqrt{2}\sigma}.
\labell{Rasymlarge}
\eeq
This is precisely the leading order solution found above with
the universal behavior: $R\sim\sigma^{-1}$.
However, for large $R$ the equation becomes
$R'=\mp4R^4/(c\lambda^2)$, with the solution
\beq\labell{Rasymeq}
R(\sigma)\simeq\left({12\sigma\over c\lambda^2}\right)^{-1/3},
\eeq
which is precisely the harmonic behavior that we anticipated for a 
D5-brane to appear at $\sigma=0$. The cross-over between the universal and
harmonic expansion of the funnel
occurs when the two terms under the square
root are comparable, {\it i.e.},
\beq
R_c\sim (c\lambda^2/2)^{1/4}=(2\pi^2c)^{1/4}\ls.
\eeq
Note that for large $c$ (and hence large $N$), this is a macroscopic
distance scale, \ie $R_c\gg\ls$.

If this configuration indeed corresponds to a D5-brane then the fuzzy funnel 
must act as the source for correct RR fields. One can easily show that the 
following term in the non-abelian Wess-Zumino action\cite{dielec,watiprep}
gives,
\beqa
-{\lambda^2\mu_1\over2}\int\STr\,P\left[(\hi_\Phi 
\hi_\Phi)^2 C^{(6)}\right]&=&
-{\lambda^3\mu_1\over2}\,\int d\sigma d\tau\,
C^{(6)}_{012345}\STr(\epsilon^{ijklm}\Phi^i \Phi^j\Phi^k\Phi^l
\partial_\sigma\Phi^m)\nonumber\\
&=&\pm{6N(n+2)\over c^{3/2}}\mu_5\int d\tau dR\ \Omega_4 R^4
\, C^{(6)}_{012345},
\labell{sourse}
\eeqa
where $\mu_5=\mu_1/(2\pi\ls)^4$ and 
$\Omega_4=8\pi^2/3$ is the area of a unit four-sphere. 
This is precisely the D5-source
term we expect. Further, the number of D5-branes is given by
\beq
{6N(n+2)\over c^{3/2}}={(n+1)(n+2)^2(n+3)\over n^{3/2}(n+4)^{3/2}}\simeq n,
\eeq
for  large $N$. Thus, in this limit, the funnel appears to
expand into $n$ D5-branes. One may also calculate the energy of this system
to be,
\beqa
E&=&NT_1\int_0^\infty d\sigma\,[1+4R^4/(c\lambda^2)]^2\nonumber\\
&=&NT_1\int_0^\infty d\sigma+{6N\over c} T_5\int_0^\infty \Omega_4 R^4 dR
+NT_1\int_0^\infty dR - \Delta E,
\labell{energy}
\eeqa
where $T_5=T_1/(2\pi\ls)^4$. Where the first term corresponds to the energy of the $N$ 
semi-infinite D-strings that we started with and the second (in the large $N$ limit) to
$n$ orthogonal D5-branes spanning the $X^{1,2,3,4,5}$ hyperplane.
As this is not a supersymmetric configuration there are other 
contributions to the energy. The third term in eqn.~\reef{energy}
seems to correspond to $N$ 
semi-infinite D-strings extending radially
{\it inside} the D5-brane. The fourth term is a finite binding energy which can
be evaluated numerically to be: $\Delta E\approx 1.0102 N c^{1/4}T_1 \ls$.

We now turn to a discussion of this new brane intersection in terms the world volume theory
of the D5-branes. The analysis of refs.~\cite{strominger,semenoff}
indicates that $N$ D-strings may terminate on a collection of $n$ 
D5-branes so long as they act as a source for second Chern class, \ie~
$\int_{S^4} \Tr F\wedge F = 8\pi^2 N$ for any $S^4$ surrounding 
the end point of the D-strings. We note that in this case producing a smooth
resolution of the D-string endpoint requires a non-abelian gauge field 
configuration and therefore $n > 1$. The action for $n$ D5-branes reduces to
\beq
S_5=-T_5\int d^6\!\sigma\ \STr\sqrt{-\det\left(G_{ab}+
\lambda^2\partial_a\phi\partial_b\phi+\lambda F_{ab}\right)}
\labell{D5action}
\eeq
if we excite only one $U(1)$ scalar, as well as the nonabelian gauge field.
In order to construct a D-string spike solution to this theory we proceed in analogy to the 
bion spike in the D3-brane theory. We work in spherical polar coordinates in the D5-brane 
world volume with a radius $r$ and angles $\alpha^i, i=1,\ldots ,4$ 
with metric,
\beq
ds^2=G_{ab}d\sigma^a d\sigma^b=-dt^2+dr^2+r^2 g_{ij}d\alpha^i d\alpha^j,
\eeq
where $g_{ij}$ is the metric on a four-sphere with unit radius,
\beq
g_{ij}={\rm diag}[1,\ \sin^2(\alpha^1),\ \sin^2(\alpha^1)\,\sin^2(\alpha^2),\ 
\sin^2(\alpha^1)\,\sin^2(\alpha^2)\,\sin^2(\alpha^3)]\ .
\labell{jdsuni}
\eeq
Now we look for bion spike solutions with a ``nearly spherically symmetric''
ansatz: The scalar is only a function of the 
radius, \ie $\phi=\phi(r)$. 
For the gauge field, we require that $A_r=0$ while the angular components
are independent of $r$, \ie, $A_{\alpha^i}=A_{\alpha^i}(\alpha^j)$.
Examining the full equations of motion shows that this is a consistent ansatz.
For the non-vanishing components of the field strength, we
introduce the convenient notation: $\tF_{ij}\equiv\lambda
F_{\alpha^i\alpha^j}$.

The gauge field equation of motion can be written
\beq
D_i\left[\sqrt{g}{r^4\tF^{ij}+{1\over4}\dtF^{ij}\tF_{kl}\dtF^{kl}
\over
\sqrt{r^8+{1\over2}r^4\tF_{ij}\tF^{ij}+{1\over16}(\tF_{ij}\dtF^{ij})^2}}
\right]=0,
\eeq
It is easy to see that if we choose a self-dual field strength
$\tF_{ij}=\dtF_{ij}$, the above equation reduces to the usual
Yang-Mills equations, which are then automatically satisfied due to the
Bianchi identity. Further the scalar equation of motion implies 
that $\tF_{ij}\dtF^{ij}$ be independent of
the angles, \ie~ we are considering homogeneous instantons on the four-sphere. 
The scalar equation of motion reduces to,
\beq
\lambda^2(\phi')^2={1\over\tB^2({n r^4\over\lambda^2}+{3\over2}N)^2-1}.
\labell{skale}
\eeq
where $\tB$ is a dimensionless integration constant.
To compare with the radial profile (\ref{Rpeq}) 
found in the D-string description, we identify the physical
transverse distance as $\sigma=\lambda\phi$, and equate the radii
$r=R$. Then we see the form of the two equations agrees provided
that we set $\tB=2/3N$. Complete agreement of the equations requires
the coefficients $4/c$ and $2n/(3N)$ to be equal, and in fact, this
equality is achieved in the large $N$ limit. Hence in this limit we have
complete agreement for the geometry determined by
the two dual approaches!

For the spherically symmetric spike describing semi-infinite D-strings,
the energy  is easily evaluated to be
\beqa
E&=&T_5\int\sqrt{g}\,d^4\alpha\, dr\ \sqrt{1+\lambda^2(\phi')^2}
\,(nr^4+{3\over2}N\lambda^2)\nonumber\\
&=&N T_1\int d\sigma+n T_5\int \Omega_4 r^4dr+NT_1\int dr-\Delta E,
\labell{D5en}
\eeqa
with $\Delta E\approx1.0102 N(6N/n)^{1/4}T_1 \ls$. Again we have full
agreement with the D-string result (\ref{energy}) in the limit of large $N$. 
Note that in analogy with the D3$\perp$D1 analysis, the above
result includes a contribution of precisely $n$ D5-branes,
while the D-string calculations yield $1/n$ corrections to the
coefficient of this term.

The one point we have not yet addressed is finding an actual gauge field
solution corresponding to a homogeneous instanton on the 
four-sphere. The ADHM construction reduces the problem
of finding instanton configurations on $\bbR^4$ to an explicit algebraic
procedure --- see, \eg ref.~\cite{atiyah}. Due to conformal invariance
of the Yang-Mills system, instanton solutions on the four-sphere can then be
produced with the usual stereographic projection of $S^4$ onto $\bbR^4$.
For instanton number $N=1$ (or --1) and gauge group $SU(2)$, one finds that
the unit size instanton located at the origin of $\bbR^4$ projects to a 
homogeneous instanton configuration on the four-sphere.
This configuration is described in some detail in ref.~\cite{nonabel} for the
interested reader. Replacing the fundamental $SU(2)$ generators by
an $n\times n$ representation allows us to embed this homogeneous solution
in an $SU(n)$ gauge theory. 
The instanton number of the resulting $SU(n)$ gauge
field is maximized by choosing the irreducible representation, giving 
\beqa
N&=&{1\over 8\pi^2}\int_{S^4}\Tr\, F\wedge F={n(n^2-1)\over6}.
\labell{suninst}
\eeqa
Now a key result is that
using a theorem by H.C.~Wang\cite{wang}, one may prove that this is the {\it
homogeneous} instanton configuration on the four-sphere with the largest
possible value of the second Chern class --- see discussion in ref.~\cite{nonabel}.
Note that for large $n$, the upper bound 
(\ref{suninst}) on $N$ agrees exactly
with the mysterious restriction (\ref{Nnrel}) that appears in the
construction of the fuzzy funnel! While this agreement is a remarkable
success of the duality between the D-string and D5-brane descriptions,
it appears that this restriction arises from our use of a spherically
symmetric ansatz. Presumably the restriction can be evaded by
considering more general field configurations, although as a practical matter
the analysis becomes much more difficult.

In this section we have examined the orthogonal intersection of a collection
of $N$ D1-branes ending on a set of $n$ D5-branes.
We have done so within the world volume theories of both the D1-branes and 
the D5-branes. The two constructions agree exactly in the large $N$ limit and
provide dual, complementary pictures of this configuration. As before, one
can verify that this agreement arises in the large $N$ regime because there
is a large interval, $\ls^2\ll r^2\ll n\ls^2$, in which both theories give
reliable results. In analogy with the D3$\perp$D1-system in the
previous section, the theory on the D1-branes involving the fuzzy four-sphere
is valid far out along the spike while near the
D5-branes the configuration is best described by the instantonic configuration
in the D5-brane world volume theory. 

\section{Discussion}

We have discussed two dual descriptions of both the D3$\perp$D1- and
the D5$\perp$D1-systems. At least in a large $N$ regime, the dual theories
yield good agreement for the energy, the RR couplings and the 
geometry of brane configurations. While one may regard the two different types
of D-branes as giving dual descriptions of the same intersection, we stress
that our analysis has been limited to the {\it low energy} world-volume theory
in each case. Hence in neither case do we have a complete description
as in both theories the configurations of interest contain singularities
which are not resolved by the low energy action.
Our analysis therefore yields two complimentary descriptions of the
D3$\perp$D1- and D5$\perp$D1-intersections. In the region far away from
the D3- or D5-branes, the
fuzzy funnel construction of the D1-brane theory is reliable,
while nearby, the bion picture of the D3- or D5-brane theory is trustworthy. 
Still a careful analysis\cite{bioncore,nonabel} shows that in the large
$N$ regime, the low energy descriptions of the two dual theories are both
reliable over a large interval of the intersection geometry. This overlap
can then be understood as the source of the strong agreement in this regime.

One can go beyond the static configurations to analyse small perturbations
propagating in these brane intersections\cite{waves,K/T,bioncore,nonabel}.
However, once again, one must be careful to reconsider the regime in
which these linearized perturbation solutions are valid. In particular,
ref.~\cite{K/T} claims there is no mechanism to suppress
the propagation of high angular momentum modes at the bion core, but
this conclusion is not valid since their linearized analysis breaks down
far from the core for these modes\cite{bioncore}.
Further the dual D1-brane description
involving a fuzzy funnel makes clear that there are only $N^2$ angular
momentum modes in the low energy spectrum. (See ref.~\cite{bioncore} for
the full details.) While only a partial analysis of the linearized fluctuations
for the D5$\perp$D1-intersection has been made\cite{nonabel}, it seems that
the fuzzy four-sphere captures oscillations of both the
geometry and the internal instantonic degrees of freedom.

There are many other extensions of the present constructions that might be
considered. For instance, one might construct fuzzy funnels with other
noncommutative geometries such as those appearing in ref.~\cite{coset}.
For example, the coset $SU(3)/U(2)$
would seem to yield a brane intersection of D-strings with D5-branes
with a more exotic five-brane geometry.
Another interesting intersection is the
D7$\perp$D1-system. In this case, the configuration is supersymmetric
and so the world-volume field theory configurations may be
more reliable. In the D7-brane theory, one would have to consider
gauge field configurations on the six-sphere with nonvanishing third
Chern character. The dual D-string funnel would now involve 
a fuzzy $S^6$. 

\section*{Acknowledgments}
This research was supported by NSERC of Canada and Fonds FCAR du Qu\'ebec.
RCM would like to thank the organizers of Strings 2001 for the opportunity to
speak at their magnificent conference.

\end{document}